# Strain-controlled responsiveness of slave half-doped manganite $La_{0.5}Sr_{0.5}MnO_3$ layers inserted in $BaTiO_3$ ferroelectric tunnel junctions.

Greta Radaelli[1,2,*] Diego Gutiérrez[1], Mengdi Qian[1], Ignasi Fina[1], Florencio Sánchez[1], Lorenzo Baldrati[2], Jakoba Heidler[3], Cinthia Piamonteze[3], Riccardo Bertacco[2] and Josep Fontcuberta[1,*]

[1]Institut de Ciència de Materials de Barcelona (ICMAB-CSIC), Campus UAB, Bellaterra 08193, Catalonia, Spain.

[2]LNESS Center - Dipartimento di Fisica del Politecnico di Milano, Como 22100, Italy

[3]Swiss Light Source, Paul Scherrer Institut, 5232 Villigen-PSI, Switzerland

*E-mail: greta.radaelli@iit.it; fontcuberta@icmab.cat.

## Abstract

Insertion of layers displaying field-induced metal-to-insulator (M/I) transition in ferroelectric tunnel junctions (FTJs) has received attention as a potentially useful way to enlarge junction tunnel electroresistance (TER). Half-doped manganites being at the verge of metal-insulator character are thus good candidates to be slave layers in FTJs. However, the phase diagram of these oxides is extremely sensitive to strain and thus can be radically different when integrated in epitaxial FTJs. Here we report a systematic study of large-area (A = 4 to 100 μm$^2$) Pt/$La_{0.5}Sr_{0.5}MnO_3$/$BaTiO_3$/$La_{0.7}Sr_{0.3}MnO_3$ (Pt/HD/BTO/LSMO) FTJs, having different thicknesses of the ferroelectric (2-3nm) and HD layers (1-2nm), grown on substrates imposing either tensile ($SrTiO_3$) or compressive ($LaAlO_3$) strains. Room-temperature electric characterization of the FTJs shows polarization-controlled ON/ OFF states. Clear evidences of field-induced M/I transition (difference between junction resistance in OFF and ON state is increased of more than one order of magnitude) are observed in junctions prepared on $SrTiO_3$ but the HD layer is generally metallic on $LaAlO_3$. Moreover, the M/I transition is only confined in an interfacial layer of the slave film thus entailing an overall reduction of TER. The orderly results reported here give some hints towards selection of HD materials and substrates for optimal FTJ responsiveness.

# 1. Introduction

Ferroelectric tunnel junctions (FTJs) have recently aroused significant interest due to their fascinating properties useful for applications in nanoelectronics, spintronics and data storage.[1,2,3] A typical FTJ consists of two metal electrodes separated by a nanometer-thick ferroelectric barrier layer, which allows electron tunneling through it. The key property of the FTJ is tunneling electroresistance (TER) – a change in the FTJ's electrical resistance upon reversal of ferroelectric polarization (P).[4,5,6] Based on simple models it was predicted that the TER effect could be as large as several orders of magnitude due to the change in the tunneling potential barrier.[5,6,7] Successful demonstrations of the TER effect in patterned junctions suitable for device applications have been achieved.[8,9,10,11,12] These experimental results provided a proof of concept for the FTJ and demonstrated the possibility for thin-film ferroelectrics to be used as a nanoscale barrier in the devices that can store binary information.

The key parameters controlling the TER ratio in FTJs are the polarization of the ferroelectric barrier layer and the screening length in the adjacent electrodes. It is imperative to realize a ferroelectric barrier with a large and stable remnant polarization for both polarization states. It is also important to achieve sufficient asymmetry in the screening lengths of the two metallic electrodes, which largely controls the magnitude of the TER ratio.[7] Note, however, that too large screening length may be detrimental for polarization stability. Optimal parameters controlling FTJ properties may be obtained by varying ferroelectric and electrode materials,[13] by interface engineering,[14] and/or by the epitaxial strain control.[15] Theoretical[16] and experimental[17] works have shown that growth of compressively strained ferroelectric films allows enhancing the polarization magnitude and aligning it normal to the interface. However, epitaxial strain imposed on the entire FTJ during epitaxial growth may also change electronic and transport properties of the electrode materials. This effect is expected to be especially notable for FTJs based on electrode materials whose properties are strongly sensitive to epitaxial strain. Such an effect may enhance or reduce the TER, depending on changes in the electronic properties of the interface, which control polarization charge screening and hence the effective tunneling barrier height. Indeed, it has been recently demonstrated that the TER of FTJs using $SrRuO_3$ electrodes can be changed up to a factor of 2 depending on the strain state of the $SrRuO_3$ bottom electrode.[18] It was also proposed that TER in FTJs could be more radically enhanced if a metal-to-insulator (M/I) phase transition in one of the electrodes can be driven by the modulation of its carrier density through ferroelectric polarization switching.[19] The resulting change of the effective barrier thickness strongly impacts the TER due to the exponential dependence of tunneling conductance on this parameter.

Half-doped manganites, such as $La_{0.5}A_{0.5}MnO_3$ (A = Ca, Sr or Ba), are excellent model systems to explore the above-described effects because their electronic and magnetic phases are highly susceptible to small changes in doping and strain.[20,21] Field effects driven by the ferroelectric polarization of an adjacent ferroelectric layer can change the doping in the manganite and cause a M/I transition. If used in a FTJ, this mechanism could result in a change of the effective thickness of the barrier layer and therefore dramatically impact the TER. This property makes half-doped manganites ideal materials to be integrated in reconfigurable tunnel junctions with potentially large electroresistance.[22,23]

We recently studied room-temperature tunneling electroresistance in $Pt/BaTiO_3/La_{0.7}Sr_{0.3}MnO_3$ (Pt/BTO/LSMO) ferroelectric tunnel junctions grown on $SrTiO_3$ (STO) substrates and reported room-temperature record value for patterned micron-size devices. It was also discovered that upon polarization switching the barrier width was not constant but was slightly modified thus indicating significant charge redistribution in the heterojunction which contributed to the measured TER.[12] It follows that even larger effects could be expected in presence of adjacent layers displaying ferroelectric polarization-induced M/I transitions. Although half-doped manganites are suitable candidates, these oxides have an extreme sensitivity to strain that largely modify their orbital and magnetic ground state,[20] by prompting either insulating or bidimensional metallic conductivity rather than the bulk 3D metallic ground state. Therefore, understanding the role of inserted half-doped manganite layers as slave electrodes in FTJs demands detailed exploration of the response of ultrathin (few unit cells) layers when grown on the ferroelectric layer and their response to epitaxial strain. Whereas the phase diagram of compressively or tensile strained half-doped films (some tens nanometers) is available,[20] this is not the case of ultrathin half-doped layers, as required in FTJs.

The objective of the present study was to find out how a thin layer of half-doped manganite $La_{0.5}Sr_{0.5}MnO_3$ (HD) behaves when incorporated in BTO-based FTJs (Pt/HD/BTO/LSMO, in brief HD-FTJs) and if this could result in an increase of TER. HD-FTJs have been grown both on STO substrates imposing tensile strain and $LaAlO_3$ (LAO) substrates imposing a compressive strain. Results show an influence of strain on the resistivity of the HD layer and also a signature of field-induced M/I transition at the HD/BTO interface, which translates in an increase of the difference between junction-resistance in OFF and ON state of more than one order of magnitude thanks to the HD presence, for junctions on STO but not in FTJs on LAO substrates. However, it will be shown that the slave HD layer introduces some non-switchable series resistance that increases the overall junction resistance thus reducing the TER. Overall, only peculiar combinations of substrate induced-strain and HD-BTO barrier thicknesses allow obtaining larger room-temperature electroresistance.

## 2. Materials and Methods

HD($t_{HD}$)/BTO($t_{BTO}$)/LSMO(30 nm) heterostructures, having HD thickness $t_{HD}$ = 1 and 2 nm and BTO thickness $t_{BTO}$ = 2 and 3 nm, were epitaxially grown by pulsed laser deposition on (001)-oriented STO and LAO single crystal substrates in a single process.[24,25] It is to be emphasized that stacks on LAO and STO substrates were grown simultaneously. Reference BTO(2, 3 and 4 nm)/LSMO//STO samples were grown with the same procedure. A complete list of the samples is given in Table 1. In the paper the following code, including information both on HD-BTO layers thicknesses and substrate, will be adopted to identify the different samples: ($t_{HD}$+$t_{BTO}$)//substrate. Top Pt layers, 20 nm thick, were deposited ex situ by sputtering. On each substrate, 36 junctions with area A ranging from 4 to 100 µm$^2$ were fabricated using photolithography and ion milling.[12] A schematic drawing of the FTJ structure is shown in Figure 1(a). A $SiO_2$ layer (not shown in Figure 1) was deposited by sputtering to create an insulating layer, thus allowing the fabrication of the Au(300nm)/Cr(7nm) contacts by e-beam evaporation and lift-off.

Electric measurements were performed in two-point geometry with a Keithley SourceMeter 2611. Positive bias indicates $V > 0$ applied to the top (Pt/Au) electrode. The ferroelectric polarization was switched by applying increasing poling voltage pulses $V_{write}$, about 0.5 s long, and subsequently reducing it back to zero. After a dwell time of few seconds, I-V characteristics were measured (-0.2 V to + 0.2 V excursions) by using triangular V(t) pulses. Resistances reported here are resistance values ($R= V/I$) at $V = 100$ mV extracted from these I-V curves. The tunnel electroresistance is defined as: TER = ($R_{OFF}$-$R_{ON}$)/$R_{ON}$, where $R_{OFF}$ is the junction resistance in the high resistance state, which, in the present case, is obtained when the BTO ferroelectric polarization is pointing towards the LSMO bottom electrode ($P_{down}$), and $R_{ON}$ is the junction resistance in the low resistance state corresponding to BTO ferroelectric polarization pointing towards the Pt top electrode ($P_{up}$) (Figure 1(b)).

Ferroelectricity was characterized by using a TFAnalyzer2000 (aixACCT Systems GmbH) applying triangular $V_{top\text{-}top}(t)$ pulses of 2 kHz. Dynamic leakage current compensation technique (DLCC) was used to minimize the impact of leakage current on the measured I-V loops [26,27]. To minimize imprint fields due to asymmetric electrode (LSMO/Pt) contributions a top-top configuration has been used in which two junctions in series are measured [28]. The voltage applied to each junction is simply as $V = V_{top\text{-}top} / \delta$, where $\delta$ is twice the BTO thickness. Further details on measurement protocols can be found elsewhere [28]. In Fig. 1(c) we show a illustrative I(V) loop recorded from two (2+3)//STO FTJs of 80 $\mu m^2$ area, connected in series in top-top configuration. I(V) collected at different frequencies are shown in SI-1. The peaks (indicated by arrows) in I(V) indicating the switching current at the coercive voltage, are well visible. We strengthen that observation of switching current peaks in tunnel junctions is by itself an experimental hallmark. In fact, so far P(V) loops measured on FTJs have been obtained using piezo-force atomic-force microscopy techniques [9,29,30].

To determine the electronic structure ion $Mn^{m+}$ ions of the HD layer prior and after Pt capping, we have used x-ray absorption spectroscopy (XAS). XAS measurements were performed on the EPFL/PSI X-Treme beamline [31] at the Swiss Light Source, Paul Scherrer Institut, Villigen, Switzerland. The spectra shown in figure 5(a) were measured in the total electron yield mode, at room temperature and normal incidence, by averaging spectra collected with x-rays linearly polarized along two perpendicular directions in the sample plane.

The technology process of simultaneous growth of films on LAO and STO and subsequent lithographic patterning steps is complex and obtaining a high yield of functioning junctions is challenging. In fact we failed to successfully process (2+2)//STO FTJs and although excellent electroresistance values were obtained for some (2+2)//LAO junctions, the data statistics precludes to derive robust conclusions on the properties of these latter junctions. However, for completeness the data for (2+2)//LAO junctions are presented in Supporting Information (SI-4).

## 3. Results

In Figures 1(d) and 1(e) we show illustrative polarization-dependent junction resistance R(V) loops obtained on a junction of sample (2+3)//STO and on a junction of sample (2+3)//LAO, respectively. In this experiment, consecutive poling (writing pulses $V_{write}$) are applied and the junction resistance is subsequently determined by measuring *I-V* curves and extracting the resistance at 100 mV. One first notice in Figures 1(d-e) that the measured resistance follows a cycle that nicely mimics that of the ferroelectric polarization P(V) loop (Fig. 1(c)), and the R(V) loops are similar to those of the reference BTO(2, 3 and 4 nm)/LSMO//STO FTJs samples, confirming that the ON/OFF resistance values are still dictated by the polarization state of the FE barrier.[12] The arrows in Figures 1(d-e) indicate the direction of the written polarization. We observe that for the up-state (polarization pointing away from the LSMO bottom electrode) the conductance is larger (ON state) than for the down-state, where the polarization points towards the LSMO electrode (OFF state), as indicated in the sketches of Figure 1(b) and as in the case of reference BTO(2, 3 and 4 nm)/LSMO//STO FTJs. Similar measurements and results are obtained for junctions of samples (1+3)//STO, (1+3)//LAO and (2+3)//LAO. From data in Figures 1(d-e), we obtain, TER ≈ 68% ((2+3)//STO) and ≈105% ((2+3)//LAO), respectively. In the case of reference BTO/LSMO//STO FTJs, systematic measurements on FTJs of different thicknesses and junction areas lead to average measured TER values of ≈ 50 (± 40) % ((0+2)//STO), 577 (± 462) % ((0+3)//STO) and $1 \times 10^4$ (± $9.4 \times 10^3$) % (0+4)//STO), respectively (in the brackets we indicate the spread of experimental values).[12] It is thus evident that both $t_{HD}$ and stress imposed by the substrate may largely affect the TER of junctions involving slave HD layers, whose properties are known to be extremely sensitive on thickness and the sign and magnitude of strain.[20] We will analyze in detail these critical issues in the following sections.

### 3.1.  ($t_{HD}$ + 3nm BTO)//STO: Varying HD barrier thickness ($t_{HD}$)

We first address the impact of introducing HD layers of various thicknesses ($t_{HD}$ = 1 and 2 nm) on the electrical properties of the junctions and their response to polarization reversal of the BTO when layers are grown on the tensile-imposing stress STO substrate. With this purpose, in Figure 2(a-c) we collect data of (1+3)//STO (orange triangles), (2+3)//STO (red circles) and the (0+3)//STO reference sample (black squares). In the insets of Fig. 2 we plot the data of the ($t_{HD}$ + $t_{BTO}$)//STO samples as a function of $t_{BTO}$. To illustrate the dispersion of experimental data, in Figure 2(a-c) we report data of several HD-FTJs measured on each sample. For the reference samples, we indicate the corresponding mean values of all junctions measured (9 for $t_{BTO}$ = 2nm, 2 for $t_{BTO}$ = 3 nm, 4 for $t_{BTO}$ = 4 nm) and the error bars represent the data dispersion.[12]

In Figure 2(a) we collect the resistance per area product (R×A) for these HD-FTJs ($t_{HD}$ = 1 and 2 nm; $t_{BTO}$ = 3 nm) and reference (0+3)//STO FTJs. Illustrative experimental poling-dependent IV curves for some of these junctions and their comparison to reference samples are shown in Supporting Information SI-2. We first note that the junction resistance largely increases when increasing $t_{HD}$ irrespectively of the polarization direction of BTO. Moreover, different values for up and down states (full and empty symbols, respectively) are observed illustrating the change of the energy barrier for tunneling upon P switching. This is clearer for the reference samples (0+$t_{BTO}$)//STO ($t_{BTO}$ = 2, 3 and 4 nm) (black squares), which also display the expected exponential increase of their junction resistance with $t_{BTO}$ (inset in Figures 2(a)). In contrast, in the used scale in Figure 2(a), the differences in R × A values for up and down states of the HD-FTJs are less apparent. In Figure 2(b) we plot the corresponding differences ($R_{OFF}$ - $R_{ON}$)×A as a function of $t_{HD}$ for a fixed $t_{BTO}$ (= 3 nm) and we compare with reference samples (black symbols) (see inset of Fig. 2(b)). Distinctive values for up and down states are measured also when an HD layer is inserted, as previously observed from Figure 2(a). Remarkably, data in Figure 2(b) show an increase of ($R_{OFF}$ - $R_{ON}$)×A with increasing $t_{HD}$ from 0 to 2 nm (see also SI-2). Finally, in Figure 2(c) we collect TER values calculated from the data in Figure 2(a) as a function of $t_{HD}$ and $t_{BTO}$ (inset). Whereas an obvious exponential increase of TER with $t_{BTO}$ is observed, in agreement with expectations, for the reference samples, it is also clear that TER display a remarkable decrease when a HD layer is inserted (see also SI-2).

From data in Figure 2, some preliminary conclusions can be drawn. The observed increase of junction resistance for <u>both</u> OFF and ON states with increasing $t_{HD}$ suggests that, in HD-FTJs grown on the tensile-stressing substrate (STO), at least a fraction of the HD layer thickness is an insulator that contributes as a series resistance in the tunnel transport across the junction. Moreover, the increase in ($R_{OFF}$ - $R_{ON}$)×A with increasing $t_{HD}$ indicates that the presence of the HD film in FTJs grown on STO enhances the change of the barrier transmittance (either barrier high or width or both) for tunneling upon P switching. However, in these HD-FTJs the TER is reduced by the presence of the additional insulating series resistance that we associate to an insulating fraction of the HD layer.

## 3.2. ($t_{HD}$ + $t_{BTO}$)//LAO

We turn now to the HD-FTJs grown on LAO substrates imposing a compressive strain. As mentioned above, HD layers are expected to display different properties due to the different ground sates of HD under tensile and compressive strain. In Figure 3(a-f) we report the results. We start by comparing the properties of identical stacks (2+3) grown on STO and LAO substrates. Next, we will inspect the differences associated to the effects of varying $t_{HD}$ for a fixed $t_{BTO}$ ($t_{BTO}$ = 3 nm): (1+3)//LAO and (2+3)//LAO. In all cases, as in Figure 2, data are compared to the reference FTJs ($t_{HD}$ = 0 nm, grown on STO). We collect the resistance per area product (R×A) measured for the up- and down-states (full and empty symbols, respectively), the difference between resistance per area product in the OFF and ON states ($R_{OFF}$ - $R_{ON}$)×A, and the TER values in Figures 3a and 3d, 3b and 3e, 3c and 3f, respectively.

### 3.2.1. Nominally identical stacks (2+3) grown on STO and LAO substrates

Data for (2+3)//LAO (blue rhombi) and (2+3)//STO (red circles) are collected in Figures 3(a-c). Representative R($V_{write}$) loops are in Fig. 1(d-e). The resistance per area product (R×A) for the up- and down-states (full and empty symbols, respectively) displays the expected distinctive $R_{ON}$ and $R_{OFF}$ values (Figure 3(a)). When comparing the (R×A) values for (2+3)//LAO (blue rhombi) and (2+3)//STO (red circles) with those of the reference (0+3)//STO sample (black squares), it is clear that (2+3)//LAO has R×A values similar to those of the reference (0+3)//STO sample but smaller than those of (2+3)//STO (see SI-3 for illustrative R($V_{write}$) raw data). Figure 3(b) evidences a decrease of ($R_{OFF}$ - $R_{ON}$)×A values for (2+3)//LAO compared to (2+3)//STO down to values comparable with, although slightly smaller than, the reference (0+3)//STO FTJs. Finally, the TER values of (2+3)//LAO (Figure 3(c)) are in average slightly larger than the ones of (2+3)//STO but lower compared to reference (0+3)//STO sample (see also SI-3).

To sum up, junction resistance values in ON and OFF states for (2+3)//LAO FTJs are similar to those of the case of (0+3)//STO reference sample but much lower than similar junctions on STO ((2+3)//STO), thus suggesting that the HD ($t_{HD}$ = 2 nm) film inserted in (2+3)//LAO FTJs is mainly metallic. Consistently, their ($R_{OFF}$ - $R_{ON}$)×A values should be comparable to those of the reference sample as the HD layer in (2+3)//LAO FTJs does not contribute to the change of the tunnel barrier properties upon P switching. This is in contrast to the case of (2+3)//STO junctions of section **3.1** where, as concluded from data in Figure 2, the HD layer is (at least a fraction of it) insulating and contributes enhancing the ($R_{OFF}$ - $R_{ON}$)×A values. The TER of (2+3)//LAO and (2+3)//STO junctions turns out to be similar because, as argued above, the larger ($R_{OFF}$ - $R_{ON}$)×A of (2+3)//STO is compensated by its larger junction resistance.

### 3.2.2. Varying $t_{HD}$ thicknesses: (1,2+3)//LAO

Figures 3(d-f) show data for (1+3)//LAO (cyan stars) and (2+3)//LAO (blue rhombi) junctions. Notice that, in these figures, superimposed data have been shifted by ±0.2 nm on the horizontal axis in order to make visualization clearer. When comparing the (R×A) values for (1+3)//LAO and (2+3)//LAO FTJs in Figure 3(d) it can be appreciated that the (R×A) and ($R_{OFF}$ - $R_{ON}$)×A values are not larger than in the reference (0+3)//STO but similar and their TER are somewhat reduced. See SI-4 for illustrative poling-dependent IV curves. These observations indicate that ultrathin (1, 2 nm) HD films inserted in ($t_{HD}$+3)//LAO FTJs are metallic. The observed reduction of ($R_{OFF}$ - $R_{ON}$) × A and TER would imply that the metallic HD layer has screening length closer to that of the bottom optimally-doped LSMO than to that of Pt.

## 3.3. Summary of main experimental observations

Data reported above indicate that, in general, the response of the ultrathin HD ($La_{0.5}Sr_{0.5}MnO_3$) layer integrated in BTO-based FTJs is primarily dependent on the strain state imposed by the substrate and also on the thicknesses of the HD and BTO layers. In Figure 4 we sketch the conclusions derived from the analysis of the junctions in sections 3.1 and 3.2 above.

From data in Figure 2(a-c) we concluded that HD layers in FTJs on STO are, at least for a fraction of their thickness, insulating and contribute positively to the change of the energy barrier for tunneling upon P switching, thus showing a signature of field-induced metal-to-insulator transition at the HD/BTO interface (labelled M/I in Figure 4(a)). However, TER variation is dominated by the presence of the remaining series resistance associated to the insulating fraction of the HD layer (labelled I in Figure 4(a)) that produces an overall reduction of TER. The role of HD in junctions on LAO substrates is found to be radically different. We have shown in Figures 3(a-f) that HD layers in FTJs on LAO are mainly metallic (this is sketched in Fig. 4(b)) when inserted in BTO-based junctions, the observed reduction in TER is attributed to the fact that the screening length in HD is closer to that of the bottom optimally-doped LSMO than in Pt.

# 4. Discussion

The starting point for rationalizing these results is the phase diagram of the HD $La_{0.5}Sr_{0.5}MnO_3$ under strain. It was predicted that $La_{0.5}Sr_{0.5}MnO_3$ under strain should evolve from an A-type antiferromagnetic (AF) ordering, involving in-plane metallic conduction for tensile strain, to a C-type insulating AF order for compressive strain.[32] These predictions were subsequently confirmed by magnetic, transport and X-ray absorption experiments on LSMO films around 40 nm thick.[20,33] The observation reported here that tensile strain (STO substrate) imprints an out-of-plane insulating character to the HD layer is in agreement with these earlier results. The observation that upon P switching the tunnel barrier transmittance contrast ($R_{OFF} - R_{ON}$)×A is larger than in the case of the bare BTO can thus be taken as a consequence of M/I transition in the HD induced by P, which results from P-induced interfacial bond and charge reconstruction. Indeed, it has been shown that the tensile strain favors charge localization at $x^2-y^2$ orbitals whereas symmetry breaking at HD/BTO interface may favor a competing $z^2$ orbital occupancy that ultimately leads to a broader effective bandwidth and larger conductivity.[34]

Turning now to FTJ's on the compressing LAO substrate, the robust metallic character of HD layers on BTO (3nm) //LAO is surprising, because on the basis of the known phase diagram of compressively strained (but thicker HD films) an insulating character was expected.[20,32]

A simple explanation for this unexpected observation can be derived from the XAS data. In Figure 5(a) we show the XAS close to the $L_{2,3}$ Mn-edge obtained from HD layers from ad-hoc prepared $La_{0.5}Sr_{0.5}MnO_3$ (2 nm)/$BaTiO_3$ (30 nm)/$La_{0.7}Sr_{0.3}MnO_3$ (30 nm)//STO(001) structures (LSMO5/BTO/LSMO//STO). The aim was to determine any possible change of electronic configuration of $Mn^{m+}$ upon capping the HD layer with a top (Pt) electrode. We include in Fig. 5(a) data collected from the bare HD film and data collected on from HD films capped with a thin (3 nm) Pt layer (capping performed *in-situ*, using the same conditions used to grow the top electrodes in the FTJs junctions, after the growth of HD and without breaking the vacuum in the growth chamber). It can be appreciated in Fig. 5(a) that when Pt is deposited on-top of the HD layer a clear feature appears in the low-energy side of the Mn absorption edge. This signals the emergence of a $Mn^{2+}$ state upon Pt capping, and thus indicates the occurrence of some reduction of the HD layer. In other words, the hole density of the HD $La_{1-x}Sr_xMnO_3$ layer is somehow reduced below the nominal (x = 0.5) value by Pt capping. The significance of this finding becomes clear when one considers the phase diagram of $La_{1-x}Sr_xMnO_3$ films under strain and doping [21]. For convenience we sketch in Fig. 5(b) the phase diagram calculated by Fang et al [21]. The horizontal lines indicate the strain imposed by the LAO (c/a > 1; compressive) or STO (c/a < 1; tensile) substrates. The vertical (black solid line) indicates the nominal doping (x=0.5). The vertical (red solid line) line indicates a phase-diagram boundary for a Pt-induced hole-depressed HD layer (we set it arbitrarily at about $x \approx 0.38$). The vertical dashed lines indicate a hole-doping variation upon polarization reversal in a neighboring BTO layer. Inspection of this figure clearly reveals that due to the asymmetric boundaries of C-AF, A-AF and FM regions as a function of strain and doping, for $x \approx 0.38$ the samples under tensile strain (HD//STO) are close to the transition between the insulating A-AF phase and the FM phase. This is fully consistent with the TER enhancement seen in Figure 3(b). However, interestingly, in case of compressive strained films (HD//LAO) the HD goes deep into the FM metallic phase, irrespectively on the P direction, so that we cannot expect any increase of the TER induced by the insertion of the HD manganite. On the contrary, an overall decrease of the TER with respect to the reference is seen (see Figure 3(b)). According to the phase diagram of Fig. 5(b), the hole-deficient (x < 0.5) HD doped manganite grown on LAO has metallic properties approaching those of the bottom electrode (x = 1/3) and consequently both should have similar screening lengths. It thus follows that TER should be reduced as experimentally observed. Finally, note that the anomalous metallic behavior of the HD electrode we derived from the analysis of the resistance of the HD FTJs on LAO can also be easily understood within the same framework.

# 5. Conclusion

In brief, two main messages emerge from the systematic study presented here. First, in general the properties of ultrathin HD layers in Pt/HD/BTO/LSMO//STO, mimic those of single HD layer, namely out-of-plane insulating behavior than produces a large increase of the junction resistance, and M-I transitions can be induced by P reversal. ($R_{OFF}$-$R_{ON}$) values can be increased up to one order of magnitude upon P reversal although in some cases TER may be reduced due to the larger junction resistance. On the contrary, unexpected metallic behavior and consequent absence of M-I transitions induced by P reversal is observed for compressively strained ultrathin HD layers (junctions on LAO substrate). From X-ray absorption experiments we have shown that the growth of the top Pt contact induces some hole depletion in the HD layer. In agreement with the known phase diagram of manganites, this has a dramatic effect on the TER, depending on strain state. A clear TER enhancement is seen only for tensile strain (STO substrate), while a TER decrease is found for hole-deficient HD under compressive strain. In the last case the manganite is pushed in the FM zone of the phase diagram, away from the boundary, thus displaying a robust metallic behaviour unaffected by the polarization reversal in BTO. Importantly, it has been observed a larger dispersion of *R*×*A* and TER values in junctions containing HD compared to those of bare BTO. This observation, together with the inherent instability of orbital order in few unit cell HD films, could be fingerprints of the known strong tendency towards phase separation in HD manganites,[35,20,33] which may constitute a bottleneck for practical use of slave HD layers in ferroelectric tunnel junctions.

# Acknowledgements

This work has been partially supported by grants from the Spanish Government (MAT2014-56063-C2-1R, MAT2015-73839-JIN and SEV-2015-0496), by the Catalan Government (2014

SGR 734) and by Fondazione Cariplo via the project MAGISTER (grant n. 2013-0726). IF acknowledges Juan de la Cierva – Incorporación postdoctoral fellowship (IJCI-2014-19102).

# Figures

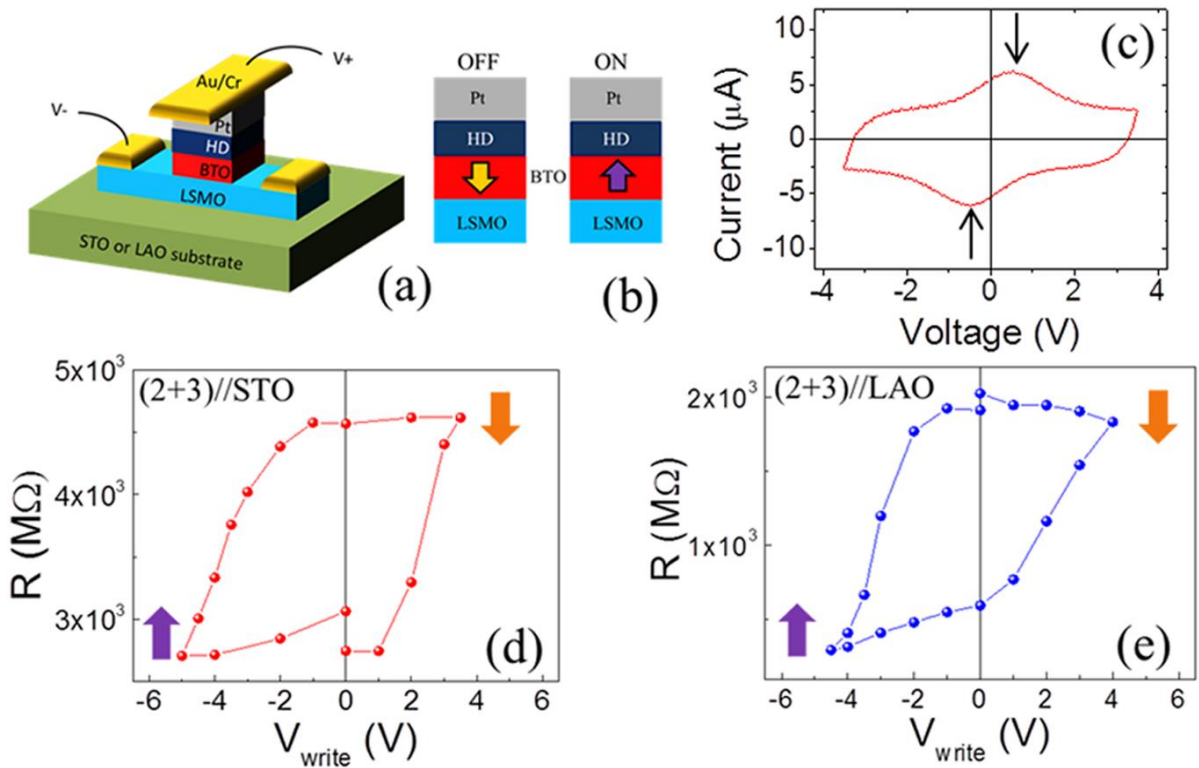

**Figure 1.** (a) Sketch of the junction structure. (b) Relative orientation of the BTO polarization (arrow) at the ON/OFF states. (c) Measured I(V) curve with DLCC method (2 kHz) on (2+3)//STO sample (4 µm$^2$) (d) and (e) Poling-dependent resistance of a representative (2+3)//STO FTJ with A = 80 µm$^2$ and (2+3)//LAO FTJ with A = 8 µm$^2$, respectively.

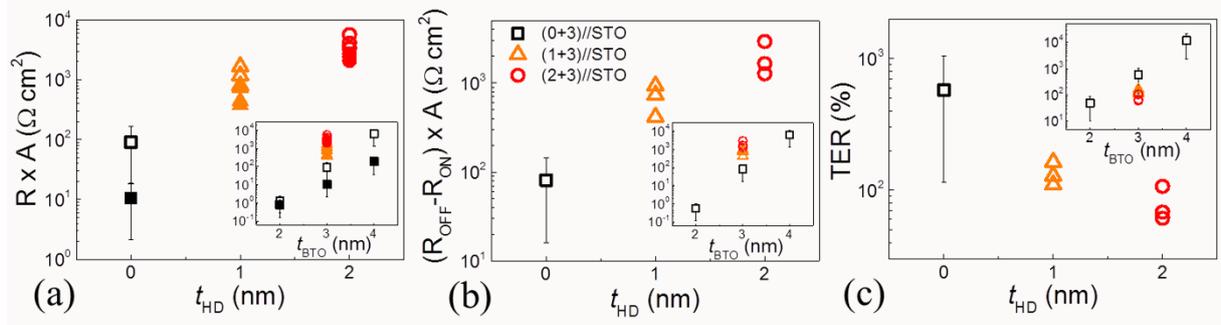

**Figure 2.** (a) Measured junction resistance per area product $R \times A$ in the OFF (empty symbols) and ON (full symbols) states, (b) difference between resistance per area product in the OFF and ON states ($R_{OFF} - R_{ON}$)$\times A$, and (c) TER, as a function of HD layer thickness: (0+3)//STO (black squares), (1+3)//STO (orange triangles) and (2+3)//STO (red circles). Black squares are average values measured on reference (0+$t_{BTO}$)//STO samples. Bars indicate the data dispersion. In the insets: data as a function of BTO layer thickness (vertical axis names as in main panels).

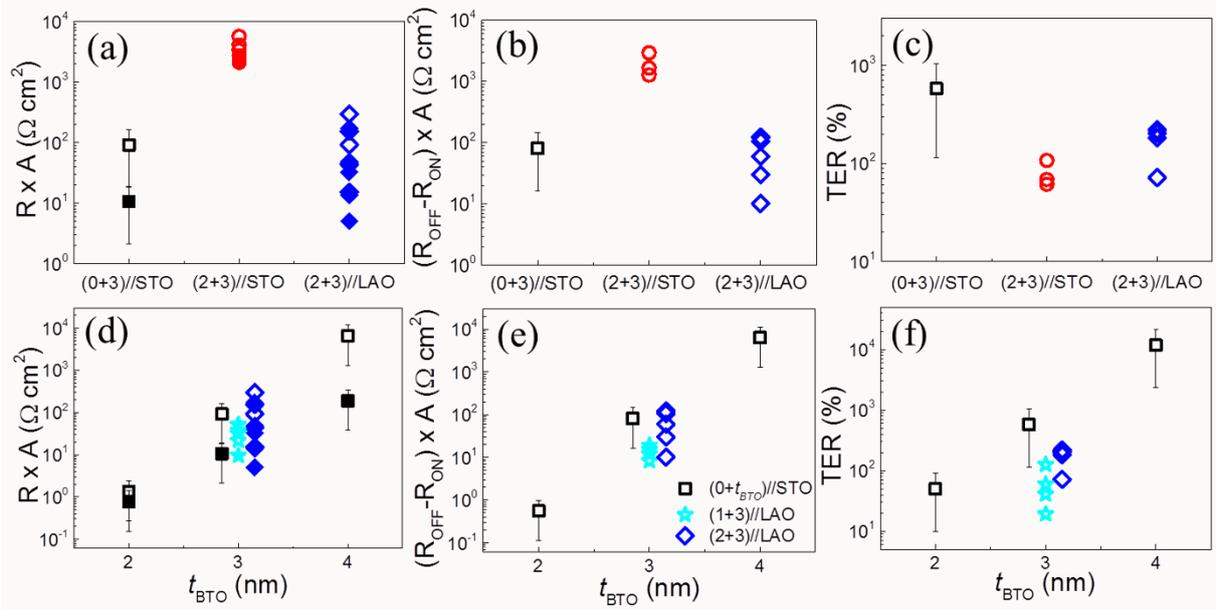

**Figure 3.** (a) Measured junction resistance per area product $R \times A$ in the OFF (empty symbols) and ON (full symbols) states; (b) difference between resistance per area product in the OFF and ON states $(R_{OFF} - R_{ON}) \times A$, and (c) TER for all measured junctions of samples (2+3)//STO (red circles) and (2+3)//LAO (blue rhombi). Black squares represent average values for reference (0+3)//STO sample. Bars indicate the data dispersion. (d) Measured junction resistance per area product $R \times A$ in the OFF (empty symbols) and ON (full symbols) states; (e) difference between resistance per area product in the OFF and ON states $(R_{OFF} - R_{ON}) \times A$, and (f) TER as a function of BTO barrier thickness for all junctions of samples (1+3)//LAO (cyan stars) and (2+3)//LAO (blue rhombi). Black squares represent average values measured on reference $(0+t_{BTO})$//STO samples. Bars indicate the data dispersion. Note that superimposed data have been shifted by ±0.2 nm on the horizontal axis in order to make visualization clearer.

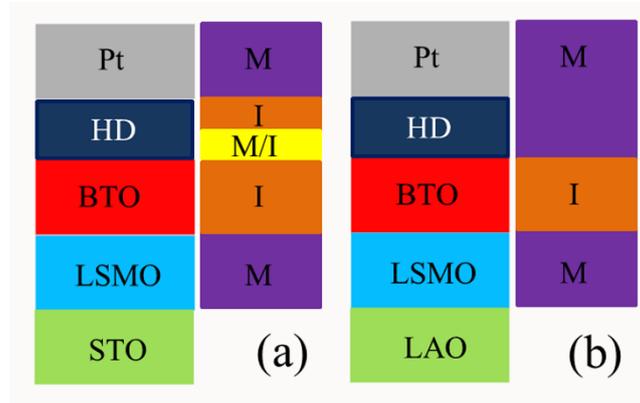

**Figure 4.** (a-b) Sketches illustrating the behavior (metallic M, insulating I or metal/insulating M/I) of the different layers of the FTJs considered in this paper. The electric character of the ultrathin HD layers (1-2 nm) inserted in FTJs involving nanometric (3 nm) BTO barriers on substrates imposing (a) tensile strain (STO) and (b) compressive strain (LAO) is depicted.

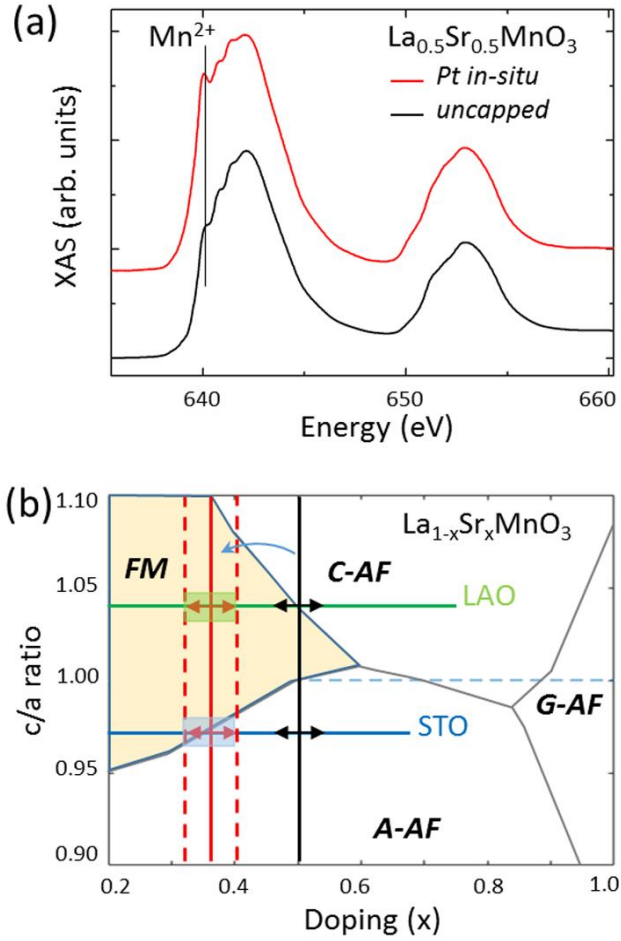

**Figure 5.** (a) X-ray absorption spectra (XAS) at L2,3 - Mn edges of $La_{0.5}Sr_{0.5}MnO_3$ (2 nm) / $BaTiO_3$ (30 nm) / $La_{0.7}Sr_{0.3}MnO_3$ (30 nm)//STO(001) films either uncapped (bottom spectrum) or in-situ capped with Pt (3nm) (top spectrum). The vertical line signals the position of an emerging $Mn^{2+}$ line most visible in the Pt-capped layer. (b) Phase diagram of strained $La_{1-x}Sr_xMnO_3$ manganites (adapted from Fang et al (Ref 21)). The horizontal lines labeled LAO and STO indicate the strain induced by LAO substrates on $La_{1-x}Sr_xMnO_3$ film. The vertical black solid line at x = 0.5 indicates the nominal hole doping (x = 0.5) of HD manganites. The vertical (red) line indicate phase boundary for a manganite films with a smaller hole concentration. We arbitrarily set here at x ≈ 0.38. The dashed vertical lines roughly indicates the expected polarization-dependent hole depletion/enrichment.

**Table 1.** List of samples and the corresponding codes used to identify them in the paper: ($t_{HD}$+ $t_{BTO}$)//substrate.

|  | Sample 1 | Sample 2 | Sample 3 | Sample 4 | Sample 5 | References |
|---|---|---|---|---|---|---|
| Sample code | (1+3)//STO | (2+3)//STO | (1+3)//LAO | (2+3)//LAO | (2+2)//LAO | (0+ $t_{BTO}$)//STO |
| $t_{BTO}$ (nm) | 3 | 3 | 3 | 3 | 2 | 2-3-4 |
| $t_{HD}$ (nm) | 1 | 2 | 1 | 2 | 2 | 0 |
| Substrate | STO | STO | LAO | LAO | LAO | STO |

# Supporting Information: Strain-controlled responsiveness of slave half-doped manganite La$_{0.5}$Sr$_{0.5}$MnO$_3$ layers inserted in BaTiO$_3$ ferroelectric tunnel junctions.

*Greta Radaelli,[*] Diego Gutiérrez, Mengdi Qian, Ignasi Fina, Florencio Sánchez, Lorenzo Baldrati, Jakoba Heidler, Cinthia Piamonteze, Riccardo Bertacco and Josep Fontcuberta[*]*

**SUPPORTING INFORMATION SI-1**

In Figure SI-1(c) we show some I(V) curves obtained using the DLCC leakage compensation method, on (1+3)//STO (80 μm$^2$) junction shown in SI-1(a), recorded at several frequencies (1kHz, 2 kHz and 3kHz) using a top – top contact configuration sketched in Fig. SI-1(b). The switching current peak, although noticeable at all frequencies [Figure SI-1(c)], becomes more apparent when increasing frequency, and proportional to it as expected. However, the uncompensated leakage current contribution is also more relevant, thus reducing the accuracy of polarization current and switching voltage determination.

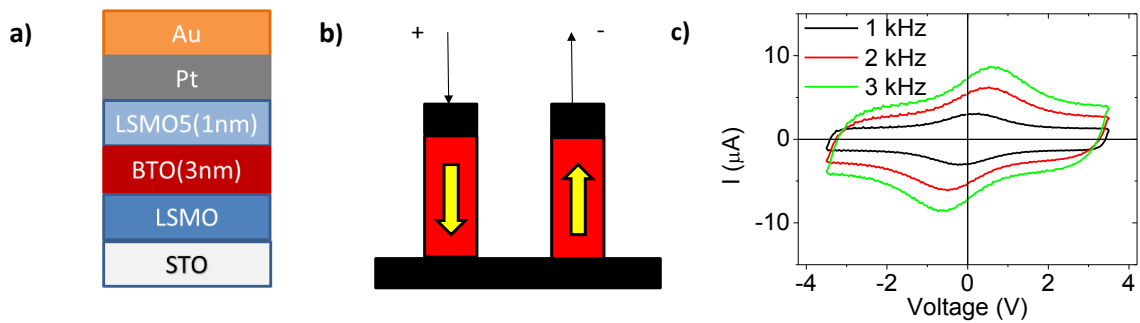

**Fig SI-1 (a)** Sketch of the 1+3)//STO (80 μm$^2$) junction, **(b)** sketch of the top-top contact configuration. **(c-**e) (I(V) loops measured with DLCC method of junctions (1+3)//STO at 1 kHz, 2 kHz and 3 kHz, respectively.

**SUPPORTING INFORMATION SI-2**

In Figures SI-2a we show the I-V curves measured after applying a writing voltage of ±4 V (solid and empty symbols) (0.5 s) on some illustrative ($t_{HD}$ + 3)//STO junctions whose data were summarized in Figure 2(a-c) of the manuscript. Distinctive I-V for up/ON and down/OFF states (solid and empty symbols respectively) are measured in the FTJs when a HD layer is inserted. It can be appreciated that insertion of the HD layer produces an increase of the junction resistance for both $R_{OFF}$ and $R_{ON}$ states (Fig. SI-2a) and an increase of ($R_{OFF}$ - $R_{ON}$)×A (Figure SI-2b, left axis). The corresponding TER values are depicted in Figure SI-2b, right axis. The reduction of TER after insertion of the HD layer results from the fact that a fraction of the HD layer remains insulation upon polarization switching thus adding a series resistance to the junction.

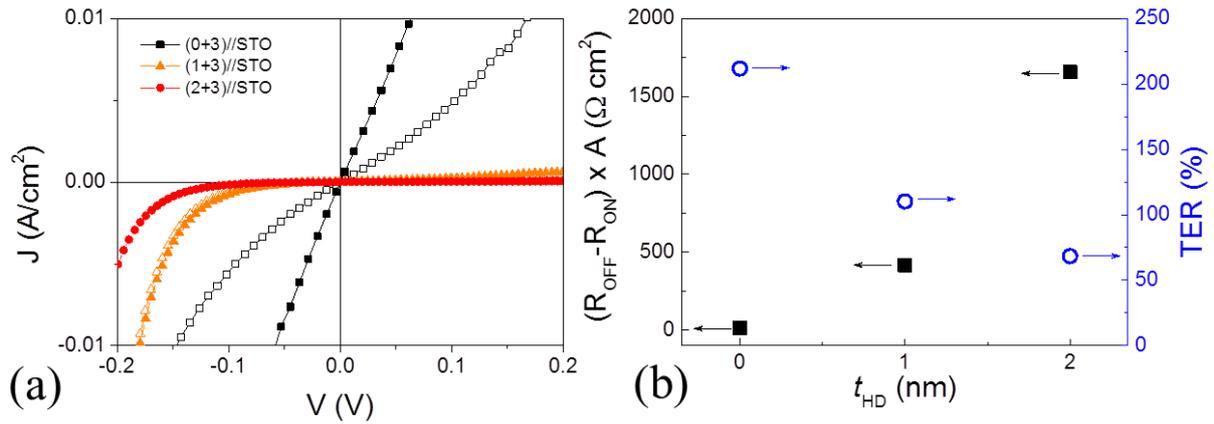

**Fig SI-2**. **(a)** Poling-dependent (-4 V and +4 V, solid and empty symbols respectively) I-V curves of a representative (0+3)//STO FTJ with A = 900 µm² (black squares), (1+3)//STO FTJ with A = 16 µm² (orange triangles) and (2+3)//STO FTJ with A = 80 µm² (red circles), respectively. **(b)** Measured difference between resistance per area product in the OFF and ON states ($R_{OFF}$ - $R_{ON}$)×A (left axis, black squares), and TER (right axis, blue circles), as a function of HD layer thickness.

**SUPPORTING INFORMATION SI-3**

In Figure SI-3a we show illustrative R($V_{write}$) loops for (2+3)//STO and (2+3)//LAO samples, whose data were summarized in Figures 3(a-c) of the manuscript. It can be clearly appreciated that the junction resistance for ON and OFF states of (2+3)//LAO is smaller than resistance values of (2+3)//STO FTJ for any polarization state. The corresponding ($R_{OFF}$ - $R_{ON}$)×A and the TER values are shown in Figure SI-3b.

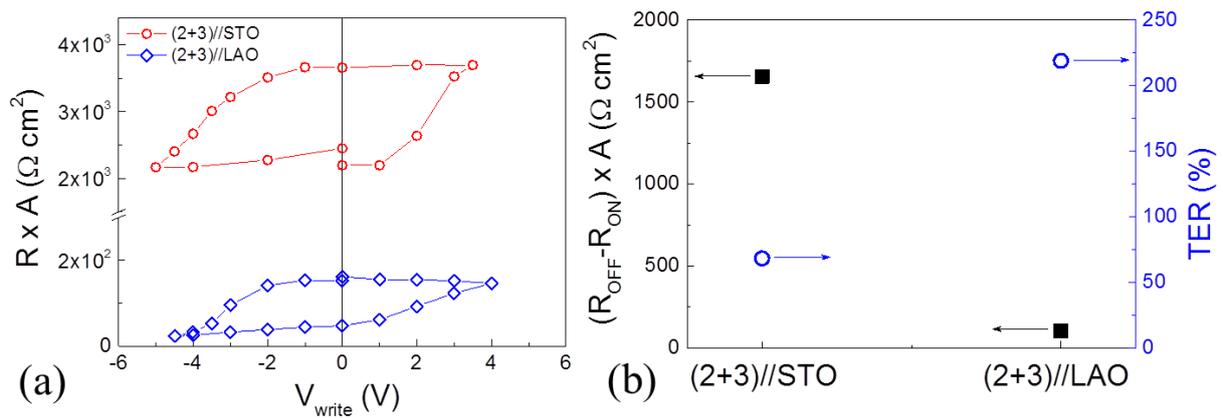

**Fig SI-3**. **(a)** Poling-dependent resistance of a (2+3)//STO FTJ with A = 80 μm² (red circles) and (2+3)//LAO FTJ with A = 8 μm² (blue rhombi), respectively. **(b)** Measured difference between resistance per area product in the OFF and ON states ($R_{OFF}$ - $R_{ON}$)×A (left axis, black squares), and TER (right axis, blue circles) for the same junctions.

**SUPPORTINGY INFORMATION SI-4**

In Figures SI-4a below we show the I-V curves measured after applying a writing voltage of ±4 V (solid and empty symbols) (10 s) of some (1+3)//LAO and (2+3)//LAO junctions and compared to the reference (0+3)//STO, whose data were summarized in Figure 3(d-f) of the manuscript. It can be appreciated in Fig. SI-4b (left axis) that the ($R_{OFF}$ - $R_{ON}$)×$A$ values are similar or slightly increased compared to those of the reference (0+3)//STO and their TER are similar or somewhat reduced (Fig. SI-4b, right axis).

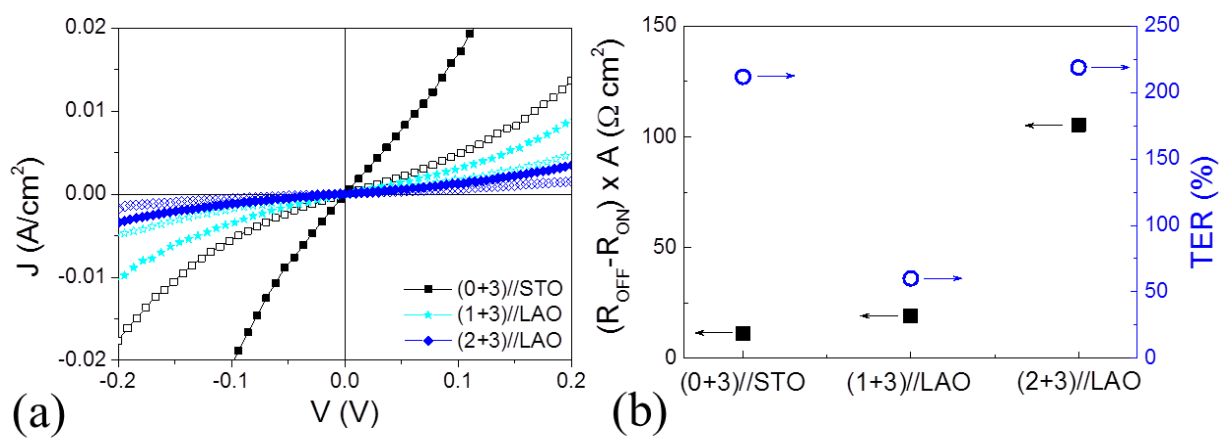

**Fig SI-4 (a)** Poling-dependent (-4 V and +4 V, solid and empty symbols respectively) I-V curves of a representative (0+3)//STO FTJ with A = 900 µm² (black squares), (1+3)//LAO FTJ with A = 30 µm² (cyan stars) and (2+3)//LAO FTJ with A = 8 µm² (red circles), respectively. **(b)** Measured difference between resistance per area product in the OFF and ON states ($R_{OFF}$ - $R_{ON}$)×$A$ (left axis, black squares), and TER (right axis, blue circles), for the three samples.